\documentclass[12]{article}

\def\s#1{#1}
\def\b#1{#1}

\newcommand{\Red}{\ }
\newcommand{\Blue}{\ }

\newcommand{\m}[1]{ {\Red $#1$} }

\newcommand{\beq}{\Red \begin{eqnarray}}
\newcommand{\eeq}{\end{eqnarray}\Blue}
\newcommand{\numeq}{\end{eqnarray}\Blue}

\newcommand{\half}{ {1\over 2} }

\newcommand{\pdr}{\partial}

\newcommand{\un}[1]{\underline{#1}}

\newcommand{\eps}{\epsilon}

\usepackage{graphics}
\input{epsf}

\newcommand{\beqs}{\begin{eqnarray}}
\newcommand{\eeqs}{\end{eqnarray}}

\def\m#1{$#1$}
\newcommand{\Em}{\em}
\def\un#1{\underline{#1}}

\def\tr{\;{\rm tr}\;}

\begin{document}

\title{\bf\large Entropy of Operator-valued Random Variables: A Variational
Principle for Large $N$ Matrix Models.}

\author{L. Akant, G. S.  Krishnaswami and S. G. Rajeev\thanks{akant,govind,rajeev@pas.rochester.edu}\\
    Department of Physics and Astronomy,\\
    University of Rochester,\\
    Rochester, New York 14627 \\
   }

 \maketitle
\begin{abstract}
We show that, in  't Hooft's large 
 \m{N}  limit,  matrix models can be formulated as a classical
theory whose equations of motion are the factorized Schwinger--Dyson equations.
 We discover an action principle for this classical theory.
 This action contains a universal term describing the entropy of the
  non-commutative probability distributions.
  We show that this entropy is a nontrivial \m{1}-cocycle of the
 non-commutative analogue of the diffeomorphism group  and derive an 
 explicit formula for it. The action principle allows us to solve matrix models
 using novel variational approximation methods; in the simple cases where
 comparisons with other methods are possible, we get reasonable agreement.
\end{abstract}

\vspace{.5in}


\vfill\eject
\tableofcontents

\vfill\eject

\section{Introduction}

There are many physical theories in which random variables which are operators
-matrices of finite or infinite order-
appear: for example,  Yang-Mills  theories, models for random surfaces and
M-theory (an  approach to a string  theory of quantum gravity). In all these
theories, the observables are functions of the matrices which are invariant
under changes of basis; in many cases - as for Yang-Mills theories-  the
invariance group is quite large  since it contains
changes of basis that depend on position. We address
the question of how to construct an effective action (probability distribution)
for these gauge invariant
observables induced by  the original probability distribution of the
matrices.

 Quantum Chromodynamics (QCD) is  the matrix model of
 greatest physical interest. QCD is the widely accepted theory of strong interactions.
It is a Yang-Mills theory with a non-abelian  gauge group \m{SU(N)}.
Thus, the microscopic degrees of freedom include  a set of  \m{N\times N}
hermitean matrices at each point of space--time: the components of the
\m{1}-form that represents the gauge field. In addition there are the quark
fields that form an \m{N}-component complex vector at each point of space--time.
The number of `colors',\m{N},  is equal to \m{3} in nature. Nevertheless,
will be useful to study the theory for an arbitrary value of \m{N}. Also, it will be convenient to regard \m{U(N)}
rather than \m{SU(N)} as the gauge group.

The microscopic degrees of freedom-quarks and gluons- do not describe the
particles that we can directly observe\cite{thooft,wittenN,makeenko}. 
Only certain bound states called hadrons-those that are
invariant under the gauge group- are observable. This phenomenon-called
confinement- is one of the deepest mysteries of theoretical physics. 

In earlier papers we have postulated that there is a self-contained theory of
color invariant observables fully equivalent to QCD at all energies and all
values of \m{N}.   We have called this theory we seek  `Quantum
HadronDynamics'\cite{2dqhd} and fully constructed it  in the case of
of  two-dimensional space--time. Also we have shown that this theory is a good
approximation to four dimensional QCD applied to Deep Inelastic Scattering: it
predicts with good accuracy 
the parton distrbutions observed in experiments\cite{quarkparton}.

Certain simplifications of the
two-dimensional theory allowed  us to eliminate all gluon ( matrix-valued ) degrees
of freedom. This  helped  us to construct two dimensional Quantum Hadron Dynamics
quite explicitly. To make further progress, it is necessary to
understand theories in which the degrees of freedom are \m{N\times N} matrices.
Before studying a full-fledged matrix field theory we need to understand how to
reformulate a theory of a finite number of matrices in terms of their
invariants. \footnote{Since we understand by now how to deal with the quark degrees of
freedom in terms of invariants, it is sufficient to consider toy models for pure
gauge theory, without vectorial degrees of freedom.} 
This is the problem we will solve in this paper.

It is well-known that matrix models simplify enormously 
 in the limit \m{N\to \infty} \cite{thooft,wittenN}. The quantum
fluctuations in the gauge invariant observables in   gauge invariant states
can be shown to be of order \m{{\hbar\over N}}. Thus, as long as we restrict to gauge invariant observables,
in the limit \m{N\to \infty} QCD must tend to some classical theory. This
classical theory cannot be Yang-Mills theory, however, since the fluctuations in
all states ( not just the gauge-invariant ones) would vanish in that limit. An
important clue to discovering Quantum Hadron Dynamics would be to study  its
 classical limit first. This is the strategy that worked in the case of two
 dimensions.

The analogue of the field equations of this `Classical Hadron Dynamics' has been
known for a long time-they are the factorized Schwinger-Dyson equations
 \cite{makeenko}. It is natural to ask if there is a variational
principle from which this equation can be derived. Finding this action principle
would be a major step forward in understanding hadronic physics: it would give
a formulation of hadron dynamics in terms of hadronic variables, entirely
independent of Yang-Mills theory. A quantization of the theory
based on this action principle would recover the corrections of order \m{1\over
N}. Moreover,we would be able to derive approximate
solutions of the large-$N$ field  equations by the variational method. 

Even after the simplifications of the large \m{N}-limit, generic matrix models
have proved to be not exactly solvable: the factorized Schwinger--Dyson  equations have
proved to be generally intractable. Diagrammatic methods have been pushed
to their limit\cite{thooft}. To make further progress, new approximation methods
 are needed- based on algebraic, geometric and  probabilistic ideas.
 Moreover, the entire theory has to be 
 reformulated in terms of manifestly \m{U(N)}-invariant observables. Thus, the
 basic symmery principle that determines the theory has to be something
 new-the gauge group acts trivially on these observables. 
 In previous papers\cite{rajeevturgut}  we had suggested that the group \m{\cal
 G} of automorphisms of a free
 algebra- the non-commutative analogue of the diffeomorphism group-
  plays this  crucial role in such a gauge invariant 
  reformulation of matrix models.
 In this paper we finally discover this manifestly gauge invariant
 formulation of finite dimensional matrix models.  We find that the
 configuration space of the theory is a coset space of \m{\cal G}- 
 justifying our earlier anticipation.

If we   restrict to   observables which are invariant under the action of
 \m{U(N)}, we should expect that the effective action should contain some kind of entropy.
The situation is analogous to that in statistical mechanics,
with the gauge invariant observables playing the role of macroscopic variables.
However, there are  an infinite number of such observables in our case.
Moreover, there is no reason to expect that the systems we are studying are in
thermal equilibrium in any sense. The entropy should be the logarithm
of the volume of the set of all hermitean matrices that yield a given set of
values for the \m{U(N)}-invariant observables. This physical idea, motivated by
Boltzmann's notions in statistical mechanics, allows us to derive an explicit
formula for entropy.

Our approach continues the point of view  in the physics literature on random
matrices \cite{wigner,mehta,thooft,wittenN,makeenko,cvitanovic,rajeevturgut,
douglas,gross-gopakumar}.
It  should not be surprising that our work has close relations to  the theory of
von- Neumann algebras-nowadays called  non-commutative  probability theory:after
all operators are just matrices of large dimension. Voiculescu \cite{voiculescu}
has another, quite remarkable, approach to  non-commutative probability
distributions. Our definition in terms of moments and the group of automorphisms is
closer in spirit to the physics literature. Also, the connection of entropy to the
cohomology of the automorphism group is not evident in that approach. A closer
connection between the mathematical and physical literature should enrich both
fields.

Although our primary motivation has been to study toy models of Yang-Mills
theory, the matrix models we study also arise in some other physical problems.
There are several recent reviews that establish these connections, so we make
only some brief comments. See e.g., \cite{guhr}.

In the language of string theory, what we seek is the  action of closed string
field theory. We solve this problem in a `toy model'- for  strings on a
 model of
space-time with a finite number of points. We find that closed string theory is
a kind of Wess-Zumino-Witten model on the coset space\footnote{\m{\cal G} is a
non-commutative analogue of the diffeomorphism group; \m{\cal SG}  is the subgroup
that preserves a non-commutative analogue of volume. See below for precise
definitions.}  \m{{\cal G}/{\cal SG}}; we discover an explicit formula for the
classical action, including a term which represents an anomaly- a nontrivial
\m{1}-cocycle. We hope that our work complements the other   approaches to
closed string field theory \cite{senreview}.

Random matrices also appear in another approach to quantum geometry
\cite{ambjorn}. Our variational method could be useful to 
approximately solve these  matrix models for Lorentzian  geometry.

Quantum chaos are often modelled by matrix models\cite{guhr}. In that context the focus is
often on universal properties that are independent of the particular choice of
the matrix model action ( which  we call \m{S} below) \cite{brezinzee}. These universal properties
are thus completely determined by the entropy. Our discovery of an explicit
formula for entropy should help in deriving such universal properties for
multi-matrix models: so far  results have been mainly about the one-matrix
model. In the current paper our focus is on the joint probability distribution,
which is definitely {\Em not} universal.

A preliminary version of this paper was presented at the MRST 2001 conference
\cite{mrst2001}.

\section{ Operator-valued Random Variables}
\def\jpd{{\rm jpd}\ }
Let \m{\xi_i,i=1\cdots M} be a collection of operator valued
random  variables. We can
assume without any loss of generality that they are hermitean operators: we can always
split any operator into its `real' and `imaginary parts.
If the \m{\xi_i} were  real-valued, we could have described
 their joint probability
distribution as a density ( more generally a measure)
on \m{R^M}. When   \m{\xi_i}  are operators that cannot be
diagonalized simultaneously , this is
not a meaningful notion; we must seek another definition for the notion of joint
probability distribution (\jpd).

The quantities of physical interest are the expectation values of functions of
the basic variables ( generators) \m{\xi_i}; the \jpd  simply provides a rule to
calculate these expectation values.
We will think of these  functions as  polynomials in the generators( more precisely
formal power series.) Thus, each random variable will be determined by a
collection of tensors \m{u=\{u^{\emptyset},u^i,u^{i_1i_2},\cdots\}} which are
the coefficients in its expansion in terms of the generators:
\beq
    u(\xi)=\sum_{m=0}^\infty u^{i_1\cdots i_m}\xi_{i_1}\cdots \xi_{i_m}.
\eeq
The constant term is just a complex number: the set of indices on it is
empty.

 If \m{u} is a polynomial, all except a finite number of the tensors will
be zero. It is inconvenient to restrict to polynomials: we would not
be able to find inverses of functions, for example, within the world of polynomials.
The opposite extreme would be to impose no restriction at all on the tensors
\m{u}: then the random variable is thought of as a formal power series. We pay a
price for this: it is no longer possible to `evaluate' the above infinite
series for any particular collection of operators \m{\xi_i}: the series may not
converge.  Nevertheless, it
makes sense to take linear combinations  and to multiply such formal power series:
\beq
[\alpha u+\beta v]^{i_1\cdots i_m}=\alpha u^{i_1\cdots i_m}+\beta v^{i_1\cdots i_m},\quad
[uv]^{i_1\cdots i_m}=\sum_{n=0}^m u^{i_1\cdots i_n}v^{i_{n+1}\cdots i_m}.
\eeq
Note that even if there are an infinite number of non-zero elements in the
tensors \m{u} or \m{v}, the sum and product is always given by finite series:
there are no issues of convergence in their definition. Thus the set of formal
power series form an associative \footnote{ This  algebra is commutative only
if the number of generators \m{M}
is one.}  algebra; this is the {\Em free algebra} \m{{\cal T}_M} on the
generators \m{\xi_i}. Note that the multiplication is just the direct product of
tensors.

As we noted above, the  joint probability distribution of the \m{\xi_i} 
is just a rule to calculate the
expectation value of an arbitrary function of these
 generators. If we  restrict to functions of \m{\xi_i} that 
are polynomials\footnote{ Not all
formal power series may have finite expectation values: the series might
diverge. This does not need to worry us: there is a sufficiently large family of
`well-behaved' random variables, the polynomials.},
such expectation values are determined by the moments
\beq
    G_{i_1i_2\cdots i_n}=<\xi_{i_1}\xi_{i_2}\cdots \xi_{i_n}>.
\eeq

If the variables commute among each other, these moments are symmetric tensors.
The most general situation that can arise in physics
is that the \m{\xi_i} satisfy no relations at all
among each other, ( in particular they dont commute)
 except the associativity of multiplication.
 In this case the moments form
 tensors with no particular symmetry property. All other associative algebras can be obtained from this `free algebra' by
 imposing relations ( i.e., quotienting by some ideal of the free algebra.)
 Such relations can be expressed as  conditions on the moments. For example,
  if \m{\xi_{i}\xi_{j}=R_{ij}^{kl}\xi_k\xi_l}, the moment tensors will satisfy
 conditions
 \beq
    G_{i_1\cdots i_ai_{a+1}\cdots i_m}=R_{i_ai_{a+1}}^{kl}
    G_{i_1\cdots i_{a-1}kli_{a+2}\cdots i_m}
 \eeq
 involving  neighboring indices.

\subsection{The Space of Paths}

Thus, in our theory, a  random variable  is a  tensor
\m{u=(u^\emptyset, u^i,u^{i_1i_2},\cdots )}. We can regard each sequence of
indices \m{I=i_1i_2\cdots i_m} as a path in the finite set \m{1,2,\cdots M}
in which the indices take their values; a tensor is then a function on this
space of paths. Now,given two paths \m{I=i_1\cdots i_m} and \m{J=j_1\cdots j_n},
 we can concatenate them: follow \m{I} after traversing \m{J}:
  \m{IJ=i_1\cdots i_mj_1\cdots j_n}. This concatenation operation is associative
  but not in general commutative; it has the empty path \m{\emptyset} as an
  identity element. There is however no inverse operation, so the concatenation
  defines the structure of a semi-group on the space of paths.

 We will use upper case latin indices to denote sequences of indices or paths.
 Repeated indices are to be  summed as usual; for example
 \beq
    u^IG_I=\sum_{m=0}^\infty u^{i_1\cdots i_m}G_{i_1\cdots i_m}.
 \eeq
Also, define \m{\delta^{I_1I_2}_I} to be  one if the paths \m{I_1} and \m{I_2} concatenate
to give the path \m{I}; and zero otherwise. \m{\bar I=i_mi_{m-1}\cdots i_1}
denotes the reverse path.
Now we can see that the direct product on tensors is just the multiplication
induced by  concatenation on the space of paths:
\beq
   [uv]^I=\delta^{I}_{I_1I_2}u^{I_1}v^{I_2}.
\eeq

A more refined notion of a path emerges if we regard the indices \m{i} as
labelling the edges of a directed graph; a sequence \m{I=i_1i_2i_3\cdots i_n} is
a path only when \m{i_1} is incident to \m{i_2}, and \m{i_2} incident with
\m{i_3} etc. The space of paths associated with a directed graph
is still a semigroup; the associative algebra induced by concatentations
 is just the algebra of functions on this semi-group. Lattice gauge theory can
 be interpreted as  a matrix model ( of unitary matrices) on a directed graph
 \footnote{ Directed graphs approximating space-time 
 are called  `lattices' in physics terminology.} 
  that approximates space-time;
  for example the cubic lattice.

 The  free algebra arises from  the graph with one
 vertex, with every edge connecting that vertex to itself; that is why every
 edge is incident to every other edge. This is the case we will mostly consider 
 in this paper. Other cases can be developed analogously.

\subsection{Non-commutative Probability Distributions }

We  {\Em define} the  `non-commutative joint probability distribution'
  of the variables \m{\xi_i} to be  a collection of tensors \m{G_{\emptyset},
 G_i,G_{i_1i_2},\cdots} satisfying the
 normalization condition
 \beq
 1=<1>,
 \eeq
 the hermiticity condition
 \beq
    G_{i_1i_2\cdots i_m}^{*}=G_{i_mi_{m-1}\cdots i_1},\quad
     \ {\rm i.e.,}\  G_{I}^{*}=G_{\bar
    I}
 \eeq
as well as the positivity condition
\beq
    \sum_{m,n=0}^\infty  G_{i_1i_2\cdots i_mj_1j_2\cdots j_n}u^{i_m\cdots
    i_1\ *}u^{j_1j_2\cdots j_n}\geq 0,\quad \ {\rm i.e.,} G_{ I
    J}u^{\bar I*}u^J\geq 0.
\eeq for any polynomial \m{u(\xi)=u^{\emptyset}+u^i\xi_i
+u^{i_1i_2}\xi_{i_1}\xi_{i_2}\cdots}. Denote by \m{{\cal P}_M} the
space of such non-commutative probability distributions. We define
the {\Em expectation value } of a polynomial
\m{v(\xi)=\sum_{m}v^{i_1i_2\cdots i_m}\xi_{i_1}\cdots \xi_{i_m}}
to be \beq <v(\xi)>=\sum_{m}v^{i_1i_2\cdots i_m}G_{i_1i_2\cdots
i_m}; \ {\rm i.e., }\ <v(\xi)>=v^IG_I \eeq

If the variables \m{\xi_i} are commutative with joint pdf \m{p(x_1,\cdots
x_n)d^M x},
\beq
    G_{i_1i_2\cdots i_n}=\int x_{i_1}\cdots x_{i_n}p(x_1,\cdots
    x_M)d^M x;
\eeq
it is clear then that the above conditions on \m{G} follow from the usual
normalization, hermiticity   and positivity conditions on \m{p(x)d^nx}.
For example, the contravariant
tensors \m{u^{\emptyset},u^i,u^{i_1i_2}} define a polynomial 
\beq
u(\xi)=\sum_{m}u^{i_1i_2\cdots
i_m}\xi_{i_1}\cdots \xi_{i_m}
\eeq
 and the quantity on the lhs of the positivity
condition above is just the expectation value of  \m{u^\dag(\xi)u(\xi)}.
In this  case, the moment tensors will be real and symmetric. Upto some
technical assumptions, the pdf \m{p(x)d^M x} is determined uniquely by the
collection of  moments \m{G_{i_1\cdots i_n}}. ( In the case of a single
variable, the reconstruction of the pdf from the moments is
 the `moment problem' of classical analysis; it  was solved at
varying levels of generality by Markov, Chebycheff etc. See the excellent book
by Akhiezer \cite{akhiezer}.)

In the non-commutative case, the pdf no longer makes sense- even
then the moments allow us  to calculate expectation values of
arbitrary polynomials. This motivates our definition.

In the  cases of interest to us in this paper, \m{G_I} is cyclically symmetric. This
corresponds to closed string theory or to glueball states of QCD. Open
strings, or mesons would require us to study moments which are not cyclically
symmetric. The theory adapts with small changes; but we wont discuss these cases
here.
\section{ Large $N$ Matrix Models}

The basic examples of such operator-valued  random variables are random matrices
of large size.

A {\Em matrix model} is a theory of random variables which are
\m{N\times N} hermitean matrices \m{A_i}, \m{i=1,\cdots M}.The matrix elements of each of the \m{A_i} are complex-valued random variables, with the joint
probability density function on \m{R^{N^2M}}
\beq
    {1\over Z(S)}e^{N\tr S(A)} d^{N^2M}A
\eeq
where \m{S(A)=\sum_{n=1}  S^{i_1i_2\cdots i_n}A_{i_1}\cdots A_{i_n}} is a
polynomial called the {\Em action}. Also, \m{Z(S)} is determined by the normalization condition:
\beq
    Z(S)=\int e^{N\tr S(A)} d^{N^2M}A.
\eeq
The expectation value of any function of the random variables is defined to be
\beq
    <f(A)>= \int f(A){1\over Z(S)}e^{N\tr S(A)} d^{N^2M}A
\eeq
The tensors \m{S^{i_1\cdots i_n}} may be chosen to be cyclically symmetric.
We assume they are such that the integrals above converge: \m{S(A)\to -\infty}
 as \m{|A|\to \infty}.
 The interesting observables are invariant under changes of bases.
 We can regard the indices \m{i=1,\cdots M} as labelling the links in some graph; then a
sequence \m{i_1\cdots i_n} is a path in this graph.
Then \m{\Phi_{i_1\cdots i_n}(A)={1\over N}\tr\left[A_{i_1}\cdots A_{i_n}\right]}
is a random variable depending on a loop in this graph.
 For the moment we will
consider every link in the graph to be  incident with every other link, so that
all sequences \m{i_1\cdots i_n} are allowed loops. In this case the loop
variables are invariant under simultaneous changes of basis in the basic
variables:
\beq
    A_i\to gA_ig^{\dag},\quad g\in U(N).
\eeq
If we choose some other
graph, a sequence of indices is a  closed loop only if the edge \m{i_2} is
adjacent to \m{i_1}, \m{i_3} is adjacent to \m{i_2} and so on.
 The invariance group
will be larger; as a result, \m{S^I} is non-zero only for closed loops \m{I}.

Given \m{S}, the moments of the loop variables satisfy  the
Schwinger-Dyson equations
\beq
     S^{J_1 i J_2} \langle \Phi_{J_1IJ_2}~\rangle +
     \delta_I^{I_1 i I_2}
    \langle~ \Phi_{I_1}\Phi_{I_2}~\rangle=0.
\eeq

This equation can be derived by considering the infinitesimal change of
variables
\beq
    [\delta_v A]_i=v_i^IA_I
\eeq
on the integral
\beq
    Z(S)=\int e^{N \tr S(A)} dA.
\eeq
The variation of a product of \m{A}'s is easy to compute:
\beq
    [\delta_v A]_J=v_i^I\delta_J^{J_1i J_2}A_{J_1}A_IA_{J_2}.
\eeq
The first term in the Schwinger-Dyson equation follows from the variation of the action under
this change. The second term is more subtle as it is the change of the measure
of integration-the divergence of the vector field \m{v_i(A)}:
\beq
    \delta_v (dA)=v_i^I{\pdr A^a_{Ib}\over \pdr A_{ib}^a}dA,\quad
    {\pdr A^a_{Ib}\over \pdr A_{ib}^a}=\delta^{I_1iI_2}_I\tr A_{I_1}\tr
    A_{I_2}.
\eeq

Returning to the Schwinger-Dyson equations, we see that they are 
  not a closed system of equations: the expectation value of the
 loop variables is related to that of the product of two loop variables.
 However,
there is a remarkable simplification as \m{N\to \infty}.

In the {\Em planar limit} \m{N\to \infty} keeping \m{S^{i_1\cdots i_n}}
fixed, the loop variables have no fluctuations\footnote{
This is called the planar  limit, since in perturbation theory, only Feynman
diagrams of planar topology contribute\cite{thooft}.
In the matrix model of
random surface theory, one is interested in another large \m{N} limit, the {\Em
double scaling limit}. Here the coupling constants \m{S^I}  have to vary as 
\m{N\to \infty} and tend to  certain critical values at a specified rate. The
fluctuations are not small in the double scaling  limit.}:
\beq
    <f_1(\Phi)f_2(\Phi)>=<f_1(\Phi)><f_2(\Phi)>+O({1\over N^2})
\eeq
where \m{f_1(\Phi),f_2(\Phi)} are polynomials of the loop variables.
This means that the probability distribution of the loop variables is
entirely determined by the expectation values ({\em moments})
\beq
    G_{i_1\cdots i_n}=\lim_{N\to \infty} <{1\over N}\tr A_{i_1}\cdots A_{i_n}>.
\eeq

Thus we get the {\Em factorized Schwinger--Dyson equations}:
\beq
     S^{J_1 i J_2} G_{J_1 I J_2} +
     \delta_I^{I_1 i I_2} G_{I_1}G_{I_2}=0.
\eeq
Since the fluctuations in the loop variables vanishes in the planar limit,
there must be some effective `classical theory' for these variables of which 
these  are the equations of motion. We now seek a variational
principle from which these equations follow.

Matrix models arise as toy models of Yang-Mills theory as well as string theory.
The cyclically symmetric indices \m{I=i_1\cdots i_n} should be interpreted as a
closed curve in space-time. The observable \m{\Phi_I} correspong to the Wilson
loop in Yang-Mills theory and to the closed string field. 

To summarize, the most important examples of non-commutative probability
distributions are large \m{N} matrix models:
\beq
    G_{i_1\cdots i_n}=\lim_{N\to \infty}\int {1\over N}\tr\left[ A_{i_1}\cdots
    A_{i_n}\right] e^{N\tr S^JA_J}{dA\over Z(S)}.
\eeq

\subsection{Example: the Wigner Distribution}

The most ubiquituous of all classical probability distributions is the Gaussian; the
non-commutative analogue of this is the Wigner distribution
\cite{wigner} ( also called the semi-circular distribution).

We begin with the simplest where  we have just one generator \m{\xi} for our algebra of random variables.
The algebra of random variables
 is then necessarily commutative and can be identified with the algebra of
formal power series in one variable.
The simplest example of a matrix-valued random variable is this:
 \m{\xi} is an \m{N\times N} hermitean matrix whose entries are mutually independent random
variables of zero mean and unit variance. More precisely,
 the average of \m{\xi^n} is,
\beq
    <\xi^n>=\int {1\over N} \tr \xi^n e^{-{N\over 2} \tr
    \xi^\dag\xi}{d^{N^2}\xi\over Z_N}.
\eeq
The normalization constant \m{Z_N} is chosen such that \m{<1>=1}.

The Wigner distribution with unit covariance is the limit as \m{N\to \infty}:
\beq
    \Gamma_n=\lim_{_N\to \infty}\int {1\over N} \tr \xi^n e^{-{N\over 2} \tr
    \xi^\dag\xi}{d^{N^2}\xi\over Z_N}.
\eeq
The factorized Schwinger--Dyson  equations reduce to the following  recursion relations for the moments:
\beq
    \Gamma_{k+1}=\sum_{m+n=k-1}\Gamma_m\Gamma_n
\eeq
Clearly the odd moments vanish and \m{\Gamma_0=1}. Set \m{\Gamma_{2k}=C_k}. Then
\beq
    C_{k+1}=\sum_{m+n=k}C_mC_n.
\eeq
The solution of the recursion relations
 give the moments in terms of the {\Em Catalan numbers}
\beq
    \Gamma_{2k}=C_k={1\over k+1}\pmatrix{2k\cr k}.
\eeq

\subsection{The Multivariable Wigner Distribution}

Let \m{K^{ij}} be  a positive matrix; i.e., such that
\beq
    K^{ij}u_{i}^*u_j\geq 0
\eeq
for all vectors \m{u^i} with zero occuring only for \m{u=0}. Then the moments of
the Wigner distribution on the generators \m{\xi_i,i=1,2,\cdots M} are
given by
\beq
\Gamma_{i_1\cdots i_n}=\lim_{N\to \infty} \int {1\over N}\tr\left[\xi_{i_1}\cdots
\xi_{i_n}\right] e^{-{N\over 2}K^{ij}\tr\xi_i\xi_j} {d^{N^2M} \xi\over
Z_N}
\eeq
( Again, \m{Z_N} is chosen such that \m{<1>=1}.)
It is obvious that the moments of odd order vanish; also, that the second moment
is
\beq
    \Gamma_{ij}=K^{-1\ ij}.
\eeq
The higher order moments are given by the recursion relations:
\beq
    \Gamma_{iI}=\Gamma_{ij}\delta^{I_1jI_2}_I \Gamma_{I_1}\Gamma_{I_2}.
\eeq
Note that each term on the rhs corresponds to a partition of the original path
\m{I} into subsequences that preserve the order. By repeated application of the
recursion the rhs can be written in
terms of a sum over all such `non-crossing partitions' into pairs.  The Catalan number
\m{C_k} is simply the number of such non-crossing partitions into pairs
of  a sequence of length \m{2k}.

Our stategy for studying more general probability distributions will be to
transform them to the Wigner distribution by a nonlinear change of variables.
Hence the group of such transformations is of importance to us. In the next
section we will study this group.

\section{ Automorphisms of the Free algebra}

The free algebra generated by \m{\xi_i} remains unchanged ( is isomorphic)
if we change to a new set of generators,
\beq
    \phi(\xi)_i=\sum_{m=0}^\infty \phi_i^{i_1\cdots i_m}\xi_{i_1}\cdots
    \xi_{i_m}
\eeq
provided that this transformation is invertible .
 We will often abbreviate \m{\xi_I=\xi_{i_1}\cdots \xi_{i_m}} so that the above
equation would be
\m{ \phi(\xi)_i=\phi_i^I\xi_I}.
The composition of  two
transformations \m{{{\psi}}} and \m{\phi} can be seen to be,
\beq
    [({{\psi}} \circ \phi)]_i^K = \sum_{n \leq |K|} \delta^K_{P_1 \cdots P_n}
    {{\psi}}_i^{j_1 \cdots j_n} \phi_{j_1}^{P_1} \cdots \phi_{j_n}^{P_n}.
\eeq
Note that the composition involves only finite series even when each of the
series \m{\phi} and \m{{\psi}} is infinite.

The inverse, $(\phi^{-1})_i(\xi) \equiv {\chi}_i(\xi)$, is
determined by the conditions
\beq
    [({\chi} \circ \phi)_i]^K = \delta^K_{P_1 \cdots P_n}
    {\chi}_i^{j_1 \cdots j_n} \phi_{j_1}^{P_1} \cdots \phi_{j_n}^{P_n} =
    \delta_i^K.
\eeq

\noindent They can be solved recursively for ${\chi}^i_J$:

\beqs
    {\chi}^i_j &=& (\phi^{-1})^i_j \cr
    {\chi}^i_{j_1 j_2} &=& - {\chi}^{k_1}_{j_1} {\chi}^{k_2}_{j_2}
    {\chi}^i_{l_1} \phi^{l_1}_{k_1 k_2} \cr
    \cdots \cr
    {\chi}^i_{j_1 \cdots j_n} &=& - \sum_{m<n} \delta^{P_1 \cdots
    P_m}_{k_1 \cdots k_n} {\chi}^{k_1}_{j_1} \cdots {\chi}^{k_n}_{j_n}
    {\chi}^i_{l_1 \cdots l_m} \phi^{l_1}_{P_1} \cdots
    \phi^{l_m}_{P_m}.
\eeqs Thus an automorphism \m{\phi} has an inverse as a formal
power series if and only if the linear term \m{\phi^i_j\xi_j} has
an inverse; i.e, if the determinant of the matrix \m{\phi^i_j} is
non-zero. The set of such automorphisms form a group \m{{\cal
G}_M=\;{\rm Aut}\;{\cal T}_M}. This group plays a crucial role in
our theory.

\subsection{Transformation of Moments under \m{{\cal G}_M}}

Given  an automorphism and a probability  distribution with moments
\m{G_I}
\beq
    \left[\phi(\xi)\right]_i=\phi_i^I\xi_I
\eeq
the expectation value of
\m{\left[\phi(\xi)\right]_{i_1}\cdots \left[\phi(\xi)\right]_{i_n}}:
\beq
\left[\phi_*G\right]_I=\sum_{n=1}^\infty {\phi}_{i_1}^{J_1}\cdots
    {\phi}_{i_n}^{J_n}G_{J_1\cdots J_n}
\eeq
We might regard these as the moments of some new probability distribution.
There is a technical problem however:
the sums may not converge, the group \m{{\cal G}_M} of formal power series includes many
transformations that may not map positive tensors to positive tensors: they may
not preserve the `measure class' of the joint probability distribution \m{G_I}.

Given some fixed  jpd ( say the Wigner distribution with unit covariance),
there is a subgroup \m{\tilde{\cal G}_M} that maps  it to probability
distributions;
this is the  open subset of \m{{\cal G}_M} defined by the inequalities\footnote{
\m{\bar I} denotes the reverse of the sequence: \m{I=i_1i_2\cdots i_n, {\bar
I}=i_ni_{n-1}\cdots i_1}}:
\beq
    \tilde{\cal G}_M=\left\{\phi\in {\cal G}_M|
    \left[\phi_*G\right]_{I_1I_2}u^{*{\bar I}_1}u^{I_2}\geq 0, \right\}.
\eeq
for all polynomials \m{u}.
Thus in the neighborhood of the identity the two groups \m{{\cal G}_M}
 and \m{\tilde{\cal G}_M} are the same; in particular,  they have
 the same Lie algebra. The point is that \m{{\cal G}_M} and \m{\tilde{\cal
 G}_M} are Lie groups under different topologies: the series \m{\phi\in \tilde{\cal
 G}_M} have to satisfy  convergence conditions implied by  the above
 inequalities.

It is  plausible that any probability distribution can be
obtained from a fixed one by some automorphism; indeed there should be many
such automorphisms. As a simple  example, the Wigner distribution with covariance \m{G_{ij}}
 can be obtained from the one with covariance \m{\delta^{ij}} by the linear
 automorphism \m{\phi^{i}_j\xi_j} provided that
 \m{G_{ij}=\sum_k\phi_i^k\phi_j^k}.Thus the space of Wigner distributions
 ( the space of positive covariance matrices) is the coset space
 \m{GL_M/O_M}.

 In the same spirit, we will regard the space of all probability distributions as
 the  coset space of the  group of automorphisms \m{\tilde{\cal G}_M/{\cal
 SG}_M}, where \m{{\cal SG}_M} is the subgroup of automorphisms
 that leave the  Wigner distribution of unit covariance invariant.  We can
 parametrize an aribitary distribution with  moments \m{G_I}
  by the transformation that relates it to the unit Wigner distribution with
  moments \m{\Gamma_I}:
  \beq
G_I=\sum_{n=1}^\infty {\phi}_{i_1}^{J_1}\cdots
    {\phi}_{i_n}^{J_n}\Gamma_{J_1\cdots J_n}
\eeq
Indeed, we will see below that all moments that differ infinitesimally from a
given one are obtained by such infinitesimal transformations. 
To rigorously justify our point of view, we must prove that ( in an appropriate
topology) the Lie group \m{\tilde{\cal G}} is the exponential of this Lie
algebra. We will not address this somewhat technical issue in  this paper.
In the next section we describe the Lie algebra in some more detail.
 \subsection{ The Lie algebra  of Derivations}

An automorphism that differs only infinitesimally from the
identity is  a derivation of the tensor algebra. Any derivation of
the free algebra is determined by its effect on the generators
\m{\xi_i}. They can be written as  linear combinations
\m{v_i^IL^i_I}, where the basis elements \m{L^i_I}  are defined by
\beq
    \left[L^i_I\xi\right]_j=\delta^i_j\xi_I.
\eeq The derivations form a Lie algebra  with commutation
relations \beq
        [L_I^i,L_J^j] = \delta_J^{J_1 i J_2} L_{J_1 I J_2}^j -
    \delta_I^{I_1 j I_2} L_{I_1 J I_2}^i.
\eeq

The change of the moments under such a
derivations is :
\beq
    \left[L_I^iG\right]_J=\delta_J^{J_1iJ_2}G_{J_1IJ_2}.
\eeq
We already encountered these infinitesimal variations in the derivation of the
Schwinger--Dyson equation.


 Let us consider the infinitisemal neighborhood of some reference distribution
 \m{\Gamma_I}--for example the unit Wigner distribution. We will assume that 
  \m{\Gamma_{I}} satisfies  the  strict positivity condition,
 \m{\Gamma_{IJ}u^{*I}u^J> 0}; i.e., that this quadratic form 
 vanishes only when the polynomial  \m{u} is identically zero. (This condition 
 is satisfied by the unit  Wigner distribution.)
 It is the analogue of the condition in classical probability theory 
  that the probability distribution does not vanish in some neighborhood of the
  origin. Then, the {\Em Hankel matrix} \m{H_{I;J}=G_{IJ}} is invertible on
  polynomials: it is an inner product.

 The infinitesimal change of  moments under a derivation \m{v=v_i^IL_I^i} is 
 \beq
\left[L_v\Gamma\right]_{k_1\cdots k_n}= v_{k_1}^I\Gamma_{Ik_2\cdots k_n}+
\;{\rm cyclic\ permutations\ in }\;(k_1\cdots k_n).
 \eeq
 Now, it is clear that the addition of   an arbitrary  infinitesimal  cyclically
 symmetric tensor \m{g_I} to \m{\Gamma_I}  can be achieved by some derivation:
 we just find some tensor \m{w_I} of which \m{g_I} is the cyclically symmetric
 part and put \m{w_{kk_1\cdots k_n}=v_k^J\Gamma_{JK}}.
 Since the Hankel matrix is invertible, we can always find such a   \m{v}.
 Thus an arbitrary infinitesimal change in
  \m{\Gamma_I} can be achieved by some \m{v_i^I}. 
  
  Indeed there will be many such derivations, differing by those that leave
  \m{\Gamma_I} invariant. The isotropy Lie algebra of \m{\Gamma_I} is defined by
  \beq
  v_{k_1}^I\Gamma_{Ik_2\cdots k_n}+
\;{\rm cyclic\ permutations\ in }\;(k_1\cdots k_n)=0.	
  \eeq
 
 We can simplify this condition for the choice where \m{\Gamma_I} is the Wigner
 distrbution.
 For \m{n=1} this is just
 \beq
    v_k^I\Gamma_I=0;
 \eeq
 for \m{n=2}, we get, using the recursion relation for Wigner moments,
\beq
v^{IjL}_{k_1}\Gamma_{k_2j}\Gamma_I\Gamma_L+k_1\leftrightarrow
k_2=0.\label{fixwigner}
\eeq
 In general, using the iterations of the Wigner recursion relation
\beq
\Gamma_{Ik_1k_2}=\Gamma_{k_1j_2}\Gamma_{k_2j_1}
\delta_I^{I_1j_1I_2j_2I_3}\Gamma_{I_1}\Gamma_{I_2}\Gamma_{I_3}
\eeq
etc.,
we get
\beqs
    \lefteqn{\Gamma_{k_1j_{n-1}}\Gamma_{k_2j_{n-2}}\cdots
    \Gamma_{k_{n-1}j_1}v_{k_n}^{I_1j_1I_2j_2\cdots j_{n-1}I_n}\Gamma_{I_1}\cdots
    \Gamma_{I_n}}\\
    & & +\;{\rm cyclic\ permutations\ in }\;(k_1\cdots k_n)=0.
\eeqs
In other words, we should lower a certain number of indices on \m{v_i^I}
using the second moment and contract  away the rest;the resulting tensor should not  have a cyclically
symmetric part. It would be interesting to find the solutions of these conditions more
explictly. We will not need this for our present work.

 \section{The Action Principle   and  Cohomology }

 We seek an  action \m{\Omega(G)} such that its variation under an
 infinitesimal automorphism
 \m{\left[L^i_IG\right]_J=\delta_J^{J_1iJ_2}G_{J_1IJ_2}}
 is the factorized SD equation:
 \beq
    L^i_I\Omega(G)=S^{J_1 i J_2} G_{J_1 I J_2} +
     \delta_I^{I_1 i I_2} G_{I_1}G_{I_2}.
 \eeq
 It is easy to identify a quantity that will give the first term:
 \beq
 L^i_I\left[S^JG_J\right]=S^{J_1 i J_2} G_{J_1 I J_2}.
 \eeq
 This term is simply the expectation value of the matrix model action.So
 \beq
    \Omega(G)=S^JG_J+\chi(G)
 \eeq
 with
 \beq
    L^i_I\chi(G)=\eta^i_I\equiv \delta_I^{I_1 i I_2} G_{I_1}G_{I_2}.
 \eeq
 This term arises from the change in the measure of integration over matrices;
 hence it is a kind of `anomaly'.

Now, in order for such a function \m{\chi(G)} to exist, the anomaly \m{\eta^i_I}
must satisfy an integrability condition 
\m{L_I^i(L_J^i\chi)-L_J^j(L_I^i\chi)=[L_I^i,L_J^j]\chi}; i.e.,
\beq
L_I^i\eta_J^j-L_J^j\eta_I^i-\delta_J^{J_1 i J_2}\eta_{J_1 I J_2}^j+\delta_I^{I_1
j I_2} \eta_{I_1 J I_2}^i=0.\label{deta}
\eeq
A straightforward but tedious calculation shows that this is  indeed satisfied.
We were not able to find a formal power series of moments  satisfying this
 condition even after many attempts.

Then we realized that, even in the case of a single matrix 
(treated in the appendix) there is no solution of the above equation!
The  condition above is in fact the statement
 that \m{\eta^i_I(G)} is a one-cocycle of the
 Lie-algebra cohomology of \m{\un{\cal G}_M} valued in the space of
 formal power series in \m{G}. (See the appendix\cite{evens}. )
 Although \m{\eta} itself is a quadratic
 polynomial in the \m{G}, there is no polynomial or even formal power series
 of which it is a variation: it repesents  a nontrivial element  of the
 cohomology of \m{\un{\cal G}} twisted by its representation on the space of
 formal power series in the moments.
 
 We need to look for \m{\chi} in some larger class 
  of functions on  the space \m{{\cal P}_M} 
 of probability
 distributions. Now, \m{{\cal P}_M=
 \widetilde{\cal G}_M/{\cal SG}_M}, a coset space of the group of automorphisms.
 We can   parametrize the moments in terms of
the automorphism that will bring them to a standard one: \m{G_I=
\left[\phi_*\Gamma\right]_I}.
 So, another way of thinking of functions on \m{{\cal P}_M} would be as functions
 on \m{\widetilde{\cal G}_M} invariant under the action\footnote{Such an idea was used 
 succesfully 
 to solve a similar problem on cohomologies \cite{ferrettirajeev}.} 
  of \m{{\cal SG}_M}. Thus, instead of power series in \m{G_I}, we will have 
    power series in the coefficients \m{\phi_i^I} 
 determining  an automorphism. 
In order to stand in for a  function on \m{{\cal P}_M}, 
such a power series would have to be invariant  under the 
subgroup \m{{\cal SG}_M}.
 
 Clearly, any power series  of \m{G_I} can be expressed as a power series 
 of the \m{\phi_i^I}: simply substitute \m{[\phi_*\Gamma]_I} for \m{G_I}.
   But there could be a 
 power series in \m{\phi} that is invariant under \m{{\cal SG}}
 and still is not  expressible as a power series in \m{G}. 
 This subtle distinction 
 is the origin of the cohomology we are discussing\footnote{We give a simple example in
 the appendix.}. We can now guess  that the quantity we seek is a function of this
 type on \m{{\cal P}_M}.

A hint is also provided by the origin of the term \m{\eta^i_I(G)} in the 
Schwinger-Dyson equations. It arises from the change of the measure of
integration over matrices under an infinitesimal, but nonlinear,  change of
variables. Thus, it should be useful to study 
 this change of measure under a finite nonlinear  change of variables-an 
 automorphism.

More precisley, let \m{\phi(A)} be a nonlinear 
 transformation \m{A_i\mapsto \phi(A)_i=\sum_{n=1}^\infty \phi_{i}^IA_I}, on the
space of hermitean \m{N\times N} matrices.
Also, let  \m{\sigma(\phi,A)={1\over N^2}\log\det J(\phi,A)}, 
where \m{J(\phi,A)} is the Jacobian determinant of \m{\phi}.

By the multiplicative property of Jacobians, we have
\beq
\sigma(\phi_1\phi_2,A)=\sigma(\phi_1,\phi_2(A))+\sigma(\phi_2,A).
\eeq

For example, \m{\sigma(\phi,A)=\log\det \phi_0} if
\m{\left[\phi(x)\right]_i=\phi_{0i}^j\xi_j} is  a linear transformation: the
Jacobian matrix is then a constant. It is convenient to 
factor out this  linear transformation
and write
\beq
\left[\phi(A)\right]_i=\phi_{0i}^j\left[{\tilde \phi}(A)_j\right],\quad
{\tilde \phi}(A)_i=A_i +\sum_{n=2}^\infty{\tilde \phi}_i^{i_1\cdots i_n}
A_{i_1}\cdots A_{i_n}.
\eeq

We will show in the appendix  that \m{\sigma(\phi,A)} can be written in terms of
the traces \m{\Phi_I={1\over N}\tr A_I}:
\beq\nonumber
\sigma(\phi,A)
&=&\log\det\phi_0+ \cr
& & \sum_{n=1}^{\infty}{(-1)^{n+1}\over n}
\tilde\phi_{i_1}^{K_1i_2L_1}\tilde\phi_{i_2}^{K_2i_3L_2}\cdots
\tilde\phi_{i_n}^{K_ni_1L_n}\Phi_{K_1\cdots K_n}\Phi_{L_n\cdots
L_1}. 
\eeq

Thus, the expectation value of \m{\sigma(\phi,A)}
with respect to some distribution can be expressed in terms of its
 moments \m{G_I=<\Phi_I>}, in the large \m{N} limit:
 \beq\nonumber
    <\sigma(\phi,A)>&=&c(\phi,G)\cr
        &=&\log \det \phi_1 + \cr
	& & \sum_{n=1}^\infty
{(-1)^{n+1}\over
n}\tilde\phi_{i_1}^{J_1i_2K_1}\tilde\phi_{i_2}^{J_2i_3K_2}\cdots
                        \tilde\phi_{i_n}^{J_ni_1K_n}G_{J_1 \cdots
            J_n}G_{K_n\cdots K_1}.
 \eeq
The above equation for \m{\sigma(\phi_1\phi_2,A)} then shows that the expectation
value \m{c(\phi,G)} satisfies the cocycle condition:
\beq
c(\phi_1\phi_2,G)=c(\phi_1,\phi_{2*}(G))+c(\phi_2,G).
\eeq
Moreover, if we now restrict to the case of infinitesimal transformations,
\m{\phi(\xi)_i=\xi_i+v_i^I\xi_I}, this \m{c(\phi,G)} reduces to \m{\eta}:
\beq
	c(\phi,G)=v_i^I\eta^i_I(G)+{\cal O}(v^2).
\eeq

Let us look at it another way: let \m{G=\phi_*\Gamma} for some reference
probability distribution \m{\Gamma}. Then the cocycle condition gives
\beq
	c(\phi_1,G)=c(\phi_1\phi,\Gamma)-c(\phi,\Gamma).
\eeq 
Choosing \m{\phi_1} to be infinitesimal gives then,
\beq
	\eta_i^I(G)=L_i^I c(\phi,\Gamma).
\eeq
Thus, we have solved our problem!:
\m{\chi(\phi)=c(\phi,\Gamma)} is a function on \m{\widetilde{\cal G}_M}
whose variation is \m{\eta}. But is it really a function on \m{{\cal
P}_M}?. 
In other words, is \m{\chi(\phi)} invariant under the right action of \m{\cal SG}?

 If
\m{\phi_{2*}\Gamma=\Gamma} the cocycle condition reduces to 
\beq
c(\phi\phi_2,\Gamma)=c(\phi,\Gamma)+c(\phi_2,\Gamma).
\eeq
We need to show that the last term is zero. 

We will only consider the case where the reference distribution \m{\Gamma} 
is the unit wignerian.
If \m{\phi_2(\xi)_i=\xi_i+v_i^I\xi_I} is infinitesimal, \m{c(\phi_2,\Gamma)} is
 just   \m{v_i^I\eta^i_I(\Gamma)=v_{i}^{JiL}\Gamma_{J}\Gamma_{L}}. 
 But, since \m{v} must leave the Wigner moments 
\m{\Gamma_I} unchanged, it must satisfy  (\ref{fixwigner}). If we contract that
equation by \m{\Gamma^{k_1k_2}} we will get \m{v_i^{JiL}\Gamma_J\Gamma_L=0}. 
Thus \m{\chi(\phi)} in invariant at least  under an infinitesimal 
\m{\phi_2\in {\cal SG}}. Within the ultrametric topology of formal power series,
the group \m{\cal SG} should be connected, so that any element can be reached by
a succession of infinitesimal transformations.

To summarize, we now have an action principle for matrix models,
 \beq\nonumber
 \lefteqn{\Omega(\phi)=\sum_{n=1}^\infty S^{i_1\cdots i_n}{\phi}_{i_1}^{J_1}\cdots
    {\phi}_{i_n}^{J_n}\Gamma_{J_1\cdots J_n}}\cr
    & & +\log\det \phi^i_{0j} +\sum_{n=1}^\infty {(-1)^{n+1}\over n}
{\tilde \phi}^{K_1i_1L_1}_{i_2}{\tilde
\phi}^{K_2i_2L_2}_{i_3}\cdots {\tilde
\phi}^{K_ni_nL_n}_{i_1}\Gamma_{K_n\cdots K_1}\Gamma_{L_1\cdots
L_n}
    .
 \eeq
 The factorized Schwinger--Dyson  equations follow from requiring that this action be
 extremal  under infinitesimal variations of  \m{\phi}.
By choosing an ansatz for \m{\phi} that depends only on  a few parameters and
 maximizing \m{\Omega} with respect to them, we can get approximate solutions
to the factorized SD equations.

\section{Entropy of Non-commutative Probability Distributions}

Whenever we restrict the set of allowed observables of a system, some 
 entropy is created: it  measures our ignorance of the variables we are not
allowed to measure. Familiar examples arise from  thermodynamics, where only a finite
number of macroscopic parameters are measured. In blackhole physics where only
the charge, mass and angular momentum of a blackhole can be measured by external
particle scattering: the interior of a blackhole is not observable to an outside
observer. 

There should be a similar 
entropy in the theory of strong interactions due to confinement: 
only `macroscopic' observables associated to hadrons are measurable by a
scattering of hadrons against each other. Quarks and gluons are observable only
in this indirect way. More precisely, only color invariant observables are
measurable.

In this paper, we have a toy model of this entropy due to confinement of gluons:
we restrict to the gauge invariant functions  
\m{\Phi_I={1\over N}\tr A_{I}}, of the matrices \m{A_1\cdots A_M}.  It turns out
that the  term \m{\chi} in the action principle above is just
the entropy caused by this restriction. 

Let \m{Q} be some  space of `microscopic' variables with a probability measure
\m{\mu}, and \m{\Phi:Q\to \bar Q} some map to a space of `macroscopic' variables. 
We can now define the volume of any subset of  \m{\bar Q}  to be the
volume of its preimage in \m{Q}: this is the induced measure \m{\bar \mu} on
\m{\bar Q}. 

In particular we can consider the volume of of the pre-image of a point \m{\bar
q\in \bar Q}. It is a measure of our ignorance of the
microscopic variables, when  \m{\bar q} is the result of  measuring
the macroscopic ones. Any monotonic function of this volume is just as good a
measure of this ignorance. The best choice 
is the logarithm of this volume, since
then it would be additive for statistically independent systems.  Let us denote
this function on \m{\Phi(Q)} by
\beq
\sigma(\bar q)=\log\left(\mu\left[\Phi^{-1}(\bar q)\right]\right).
\eeq
The average of this quantity over \m{\bar Q} is the {\Em entropy} of the
induced probability distribution \m{\bar \mu}.

Let us apply this idea to the case where the `microscopic' observable is a single
hermitean \m{N\times N} matrix \m{A}; the `macroscopic'  observable is the spectrum, the set
of eignvalues. We disregard the information in the basis in which \m{A} is
presented. We do so even if  this information is
measurable in principle; e.g., by interference experiments in quantum mechanics.
The natural measure on the space of matrices is the uniform (Lebesgue) measure
\m{dA} on \m{R^{MN^2}}. Althogh the uniform measure is not normalizable, 
the volume of the space of matrices with a given spectrum \m{\{a_1,\cdots a_N\}}
is finite \cite{mehta}. Upto a constant ( i.e., depending only on \m{N}), it is 
 \m{\prod_{1\leq i<j\leq N}(a_i-a_j)^2}. Thus
the entropy is \m{2\int_{x<y} \rho(x)\rho(y)\log|x-y|dxdy} where
\m{\rho(x)={1\over N}\sum_{i=1}^N\delta(x-a_i)}. This
expression make sense even in the limit \m{N\to \infty}: 
we get a continuous distribution of eigenvalues \m{\rho(x)}.

What is the `joint spectrum' of a collection of \m{M} hermitean matrices
\m{A_1\cdots A_M}? Clearly they cannot be simultaneously diagonalized, so a
direct definition of this concept is impossible. Now,
recall that the set of eigenvalues \m{\{a_1\cdots a_N\}} can be recovered from 
the elementary symmetric functions \m{G_n={1\over N}\sum_{i=1}^Na_i^n} as the
solutions of the algebraic equation
\beq
	{1\over N}x^N=G_1 x^{N-1}-G_2x^{N-2}+\cdots (-1)^{N-1}G_N.
\eeq
 The moments \m{G_n} for \m{n>N} are not independent: they can be
expressed as polynomials  of the lower ones. Although the set \m{\{a_1\cdots a_N\}}  is
determined by the sequence \m{G_1,\cdots G_N}, there is no explicit
algebraic formula: Galois theory shows that this is impossible for \m{N>4}.
Galois theory also shows that 
any gauge invariant polynomial  of \m{A} can be expressed as a polynomial of 
the \m{G_1,\cdots G_N}. Thus we can regard this sequence 
\m{{1\over N}\tr A,{1\over N}\tr A^2\cdots } as the spectrum of the matrix \m{A}.

The volume \m{\prod_{i<j}(a_i-a_j)^2} of the space of matrices with a given spectrum is a symmetric 
polynomial of order \m{N(N-1)\over 2} in the eigenvalues. Hence, in principle, 
it can be expressed as a polynomial in \m{G_1,\cdots G_N}, athough there does not appear to be a
simple universal formula\footnote{It is possible to get a formula 
for the volume in terms of the first \m{2N} moments.
The complication is that only the first \m{N} moments can be freely specified.
 The remaining moments are  determined by these , and
yet,  there is no algebraic formula that expresses \m{G_{N+1}\cdots G_{2N}}
 in terms of \m{G_1\cdots G_N}.}.

This point of  view suggests a generalization  to several matrices: we can define the {\Em joint
spectrum} of a collection of matrices to be the quantities \m{G_{i_1\cdots
i_n}={1\over N}\tr A_{i_1}\cdots A_{i_n}}. Again, there are relations among
these quantities when \m{I} is longer than \m{N}; but it is difficult to
characterize these relations explicitly. Nevertheless, it is meaningful
to ask for the volume (with respect to the uniform measure on \m{R^{MN^2}}) 
of the set of all matrices  with a given value for the sequence 
\m{G_{i_1 \cdots i_n}}.
Again, we will not get any explicit formula for entropy by pursuing this point of
view.

So we look for yet  another way to think of the joint spectrum of a 
collection of matrices. We can ask how the entropy of a collection of matrices
with joint spectrum \m{G_I} 
changes if we  transform them by some power series:
\beq
	A_i\mapsto \phi(A)_i=\phi_i^IA_I.
\eeq
Let \m{c(\phi,G)} be this change. Then, if we perform another transformation, 
we must have
\beq
	c(\phi\phi_2,G)=c(\phi,\phi_{2*}G)+c(\phi_2,G);
\eeq
i.e., it must be a 1-cocyle. Under infinitesimal variations, it reduces  to
\m{\eta}, since it is just the infinitesimal change in the uniform measure
\m{dA}.

In the last section we obtained this  \m{c(\phi,G)} explicitly as a formal
power series in \m{G}. It can be written as the variation
\beq
c(\phi,G)=\chi(\phi_*(G))-\chi(G)
\eeq
of some function \m{\chi} of the joint spectrum \m{G}. However this \m{\chi} is
not a formal power series in \m{G}, so we cannot get an explicit formula for it.
We can write it as an explicit formal power series in \m{\phi} which is
invariant under the action of \m{\cal SG}.

Thus we see the confluence of three apparently unrelated questions: an action
principle for the planar limit of matrix models ( our main interest), cohomology
of the automorphism of formal power series and entropy of non-commutative 
variables.

Voiculsecu has a somewhat different approach \cite{voiculescu} 
to defining the entropy of
noncommuting random variables. Upto some additive constant his definition seems
to agree with ours. 

\section{Example: Two-Matrix Models}

Let us consider a quartic multi-matrix model with action

\beq
    S(M) = - \tr [\half K^{ij} A_{ij} + {1 \over 4} g^{ijkl} A_{ijkl}].
\eeq

\noindent Our reference action is the gaussian
\footnote{In the language of non-commutative probability theory we used earlier,
 what we call the gaussian in this section is really the multivariate 
 wignerian distribution. There should  be no confusion, since the wignerian
 moments are 
 realized by a gaussian distribution of matrices.}
 $S_0(M) = - \tr
\half \delta^{i j} A_i A_j$. We are interested in estimating the
greens functions and vacuum energy in the large $N$ limit:

\beqs
    E^{\rm exact} &=& -\lim_{N \to \infty} {1 \over N^2} \log{Z \over
                    Z_0}
\eeqs

\noindent where $Z$ and $Z_0$ are partition functions for $S$ and
$S_0$. Choose the linear change of variable $A_i \to \phi_i(A) =
\phi^j_i A_j$. The variational matrix $\phi^j_i$ that maximizes
$\Omega$ determines the multi-variable Wigner distribution that
best approximates the quartic matrix model. For a linear change of
variables,

\beq
    \Omega[\phi] = \tr\log[\phi^j_i] -
        \half K^{i j} G_{i j} - {1 \over 4} g^{ijkl} G_{ijkl}.
\eeq

\noindent Here $G_{i j} = \phi^k_i \phi^k_j$ and $G_{ijkl} =
G_{ij} G_{kl} + G_{il}G_{jk}$ are the greens functions of
$S_0(\phi^{-1}(A))$. Thus, the matrix elements of $G$ may be
regarded as the variational parameters and the condition for an
extremum is

\beq
    \half K^{pq} + {1 \over 4} [g^{pqkl}G_{kl} +
    g^{ijpq}G_{ij} + g^{pjqk}G_{jk} + g^{ipql}G_{il}] = \half
    [G^{-1}]^{pq}
\eeq

\noindent This is a non-linear equation for the variational matrix
$G$, reminiscent of the self consistent equation for a mean field.
To test our variational approach, we specialize to a two matrix
model for which some exact results are known from the work of
Mehta \cite{mehta-two-matrix}.

\subsection{Mehta's Quartic Two-Matrix Model}

Consider the action

\beqs
    S(A,B) &=& -\tr [\half (A^2 + B^2 - c AB - c BA) + {g \over 4}(A^4 +
    B^4)].
\eeqs

\noindent which corresponds to the choices $K^{ij} = \pmatrix{1 &
-c \cr  -c & 1}$, $g^{1111} = g^{2222} = g$ and $g^{ijkl} =0$
otherwise. \footnote{Kazakov relates this model to the Ising model
on the collection of all planar lattices with coordination number
four \cite{Kazakov-rand-ising}.} We restrict to $|c| < 1$, where
$K^{ij}$ is a positive matrix. Since $S(A,B)=S(B,A)$ and $G_{AB} =
G_{BA}^*$ we may take

\beq
    G_{ij} = \pmatrix{\alpha & \beta \cr \beta & \alpha}
\eeq

\noindent with $\alpha, \beta$ real. For $g > 0$, $\Omega$ is
bounded above if $G_{ij}$ is positive. Its maximum occurs at
$(\alpha, \beta)$ determined by $\beta = {c \alpha \over 1+2g
\alpha}$ and

\beq
    4 g^2 \alpha^3 + 4 g \alpha^2 + (1-c^2-2g) \alpha -1 = 0.
\eeq

\noindent We must pick the real root $\alpha(g,c)$ that lies in
the physical region $\alpha \geq 0$. Thus, the gaussian ansatz
determines the vacuum energy ($E(g,c) = -\half \log{(\alpha^2 -
\beta^2)}$) and $\it all$ the greens functions (e.g. $G_{AA} =
\alpha, ~ G_{AB} = \beta, ~ G_{A^4} = 2 \alpha^2$ e.t.c)
approximately.

By contrast, only a few observables of this model have been
calculated exactly. Mehta \cite{mehta-two-matrix} \footnote{Some
other special classes of greens functions are also accessible (see
\cite{Staudacher-two-mat,douglas}).} obtains the exact
vacuum energy $E^{ex}(g,c)$ implicitly, as the solution of a
quintic equation. $G^{ex}_{AB}$ and $G^{ex}_{A^4}$ may be obtained
by differentiation. As an illustration, we compare with Mehta's
results in the weak and strong coupling regions:

\beqs
    E^{ex}(g,\half) &=& -.144 + 1.78 g - 8.74 g^2 + \cdots \cr
    G_{AB}^{ex}(g,\half) &=& {2 \over 3} - 4.74 g + 53.33 g^2 + \cdots \cr
    G_{AAAA}^{ex}(g,\half) &=& {32 \over 9} - 34.96 g + \cdots
    \cr \cr
    E^{var}(g,\half) &=& -.144 + 3.56 g - 23.7 g^2 + \cdots \cr
    G_{AB}^{var}(g,\half) &=& {2 \over 3} - 4.74 g + 48.46 g^2 + \cdots \cr
    G_{AAAA}^{var}(g,\half) &=& {32 \over 9} - 31.61 g + 368.02 g^2 +
    \cdots {\rm ~ e.t.c.}
    \cr \cr
    E^{ex}(g,c) &=& \half \log g + \half \log{3}-{3 \over 4}
                + \cdots \cr
    G_{AB}^{ex}(g,c) &\to& 0 {\rm ~as~} g \to \infty \cr
    G_{A^4}^{ex}(g,c) &=& {1 \over g} + \cdots.
    \cr \cr
    E^{var}(g,c) &=& \half \log{g} + \half \log{2} + {1 \over \sqrt{8g}}
        + {\cal O}({1 \over g}) \cr
    G_{AB}^{var}(g,c) &=& {c \over 2g} - {c \over (2g)^{3 \over 2}} + {\cal O}({1 \over g^2}) \cr
    G_{A^4}^{var}(g,c) &=& {1 \over g} - {2 \over (2g)^{3 \over 2}} + {\cal O}({1 \over
    g^2}), ~ \rm {e.t.c.}
\eeqs

We see that the gaussian variational ansatz provides a 
reasonable first approximation in both the weak and strong coupling regions.
 The gaussian variational ansatz is not good near singularities of the free
 energy (phase transitions).
 As \m{|c|\to 1^-}, the energy  \m{E^{ex}} diverges; this is not
 captured well  by the gaussian ansatz. This reinforces  our view that the gaussian
 variational ansatz is the analgoue of  mean field theory.

\subsection{Two-Matrix Model with $\tr [A,B]^2$ Interaction}

The power of our variational methods is their generality. We
present an approximate solution to a two matrix model, for which
we could find no exact results in the literature. The action we
consider is:

\beq
    S(A,B) = -\tr [{m^2 \over 2}(A^2 + B^2) + {c \over 2}(AB + BA) -
    {g \over 4} [A,B]^2].
\eeq

\noindent This is a caricature of the Yang-Mills action. A
super-symmetric version of this model is also of interest (see
 \cite{kawai-et-al}). Consider the regime where
$K^{ij} = \pmatrix{m^2 & c \cr c & m^2}$ is positive, $k = (m^4 -
c^2) \geq 0$. As before, we pick a gaussian ansatz and maximize
$\Omega$. We get $\beta = -{c \over m^2} \alpha$ and

\beqs
    G^{AA} &=& G^{BB} = \alpha = {m^2 \over 2g}
        [\sqrt{1 + {4g \over k}} - 1], \cr
    E &=& - \half \log{[{2 g + k - \sqrt{k^2 + 4 k g} \over 2
                g^2}]}.
\eeqs

\noindent All other mean field greens functions can be expressed
in terms of $\alpha$.

It is possible to improve on this gaussian variational ansatz  by using
nonlinear transformations. It is much easier to find first a  gaussian 
approximation  and then expand around it in a sort of loop expansion.
This is the analogue of the usual Goldstone methods of many body theory.
 We have performed such calculations for these multimatrix models, but we will not report
on them in this paper for the sake of brevity. The results are qualitatively the
same. In the next section ( an appendix) we will give the departures from the 
gaussian ansatz in the case of the single-matrix models.

\section{  Appendix: Group cohomology}


\s{ Given a group \m{ G} and a \m{ G}-module \m{ V}
(i.e., a representation of \m{G} on a vector space \m{V}),
we can define a cohomology theory\cite{evens}. The \m{r}-cochains are functions
\beq
    f:{ G}^r\to { V}.
\eeq
The coboundary is
\beqs
    df(g_1,g_2,\cdots g_{r+1})&=&
g_1f(g_2,\cdots g_{r+1})\cr
& & +\sum_{s=1}^r(-1)^sf(g_1,g_2,\cdots
g_{s-1},g_{s}g_{s+1},g_{s+2},\cdots g_{r+1})\cr & &
+(-1)^{r+1}f(g_1,\cdots, g_{r}). \eeqs It is straightforward to
check that \m{d^2f=0} for all \m{f}. A chain \m{c} is a {\Em
cocycle } or is {\Em closed}
 if \m{df=0}; a cocycle is {\Em exact} or
is a {\Em coboundary} if \m{b=df} for some \m{f};
 The \m{r}th cohomology of \m{G} twisted by the module \m{V}, \m{H^r(G,V)} is
 the space of closed chains modulo exact chains.
}
\b{\m{H^0(G,V)} is the space of invariant elements in \m{V}; i.e., the space
of \m{v} satisfying  \m{gv-v=0} for all \m{g\in G}.
 A \m{1}-cocycle
is a function \m{c:G\to V} satisfying
\beq
    c(g_1g_2)=g_1c(g_2)+c(g_1).
\eeq
Solutions to this equation modulo \m{1}-coboundaries (which are of the
form \m{b(g)=(g-1)v} for some \m{v\in V})  is
the  first cohomology \m{H^1(G,V)}.
}
\b{ If \m{G} acts trivially on \m{V}, a cocycle is just a homomorphism
of \m{G} to the additive group of \m{V}:
\m{c(g_1g_2)=c(g_2)+c(g_1)}. }

\s{A \m{1}-cocycle gives a way of turning  a representation on  \m{V}
into an affine action:
\beq
    (g,v)\mapsto  gv+c(g).
\eeq
If \m{c(g)} is a coboundary (i.e., \m{b(g)=(g-1)u} for some \m{u}),
this affine action is really a linear representation in disguise:
if the origin is shifted by \m{u} we can reduce it to a linear
representation. Thus the elements of \m{H^1(G,V)} describe
`true' affine actions on \m{V}.
}
\b{ For example let \m{G} be the loop group of a Lie group \m{G'}, the
space of smooth functions from the circle to \m{G'}:
\m{G=S^1G'=\{g:S^1\to G'\}}. Let \m{V=S^1\un{G'}} be the corresponding
loop of the Lie algebra \m{\un{G'}} of \m{G'}. Then there is an
obvious adjoint representation of \m{G} on \m{V}; a non-trivial
\m{1}-cocycle is \m{c(g)=gdg^{-1}}, \m{d} being the exterior
derivative on the circle:
\beq
    c(g_1g_2)=g_1[g_2d(g_2^{-1})]g_1^{-1}+g_1dg_1^{-1}={\rm ad}\;
g_1 c(g_2)+c(g_1).
\eeq
}

\section{ Appendix: A Single Random Matrix}
\b{In the special case where there is only one matrix (\m{M=1}), there is a
probability distribution on the real line \m{\rho(x)dx} such that
\beq
    G_n=\int x^n \rho(x)dx
\eeq
This follows because the \m{G_n} satisfy the positivity condition
\beq
    \sum_{m,n=0}^\infty G_{n+m}u_m^*u_n\geq 0;
\eeq
upto technical conditions, any  sequence of real numbers satisfying this
condition determine a probability distribution on the real line. ( This is the
{\Em classical moment problem} solved in the nineteenth century \cite{akhiezer}.)
}
\b{There is an advantage to transforming the factorized SD equations into 
an equation for
\m{\rho(x)}-it becomes a linear integral equation, the {\Em Mehta-Dyson}
equation\cite{mehta}:
\beq
        2{\cal P}\int_a^b {\rho(y) dy\over x-y} +S'(x)=0.
\eeq
}
\b{Moreover, the solution\cite{planarsaclay,brezinzee} to this equation can be expressed in purely algebraic
terms\footnote{
For a Laurent series \m{X(z)=\sum_{k=-\infty}^m X_k z^{k}} around infinity,
 \m{\lfloor X(z)\rfloor=\sum_{k=0}^m X_k z^k} denotes the part that is a polynomial in
\m{z}. This
is analogous to the `integer part' of a real number, which explains
the notation.
}
\beq
\rho(x)=-{1\over 2\pi}\theta(a\leq x\leq b)\surd[(x-a)(b-x)]\left\lfloor
{S'(x)\over \surd(x-a)(x-b)}\right\rfloor.
\eeq
The numbers \m{a} and \m{b}  are solutions of the algebraic equations
\beqs
    \sum_{r,s=0}[r+s+1]S_{r+s+1}
{(\half)_{r}(\half)_s\over r!s!}a^r b^{s}&=&0\cr
\sum_{r,s=0}[r+s]S_{r+s}{(\half)_{r}(\half)_s\over r!s!}a^r
b^{s}&=&2 \eeqs } \b{where \m{(\half)_r = \half (\half +1) \cdots
(\half+r-1)}. The simplest example is the case of the Wigner
distribution. It is the analogue of the Gaussian in the world of
non-commutative probability distributions. For, if we choose the
matrix elements of \m{A} to be independendent Gaussians,
\m{S(A)=-\half \tr A^2}, we get the distribution function for the
eigenvalues of \m{A} to be ( in the large \m{N} limit) \beq
    \rho_0(x)={1\over 2\pi}\surd\left[4-x^2\right]\theta(|x|<2).
\eeq
The odd moments vanish; the even moments are then given by the {\Em Catalan numbers}
\beq
G_{2k}=C_k={1\over k+1}\pmatrix{2k\cr k}.
\eeq
}

\s{The Mehta-Dyson equation follows from maximizing the `action'
\beq
    { \Omega}(\rho)=\int \rho(x)S(x)dx+{\cal P}\int
    \log|x-y|\rho(x)\rho(y)dxdy
\eeq
with respect to \m{\rho(x)}. Then generating function \m{\log Z(S)} is the maximum of
this functional over all probability distributions \m{\rho}.
}
\b{The physical meaning of the first term is clear: it is just the expectation
value of the action of the
original matrix model:
\beq
\int \rho(x)S(x)dx=\sum_n{G_n}S_n.
\eeq }
\b{The second term can be
thought of as the `entropy' which arises because we have lost the information
about the angular variables in the matrix variable: the function \m{\rho(x)} is
the density of the distribution of the eigenvalues of \m{A}. Indeed,
\m{\sum_{i\neq j}\log|a_i-a_j|} is (upto a constant depending only on \m{N})
the log of the volume of the space of all
hermitean matrices with spectrum \m{a_1,a_2\cdots a_N}.The entropy
\m{{\cal P}\int \log|x-y|\rho(x)\rho(y)dxdy} is  the large \m{N} limit
of this quantity.
}
\b{Note that the entropy is independent of the choice of the matrix model
action: it is a universal property of all one-matrix models.
}
\b{The meaning of the variational principle is now clear: we seek the
probability distribution of maximum entropy that has a given set of moments
\m{G_r} for \m{r=1,\cdots n}. The coefficients of the polynomial \m{S} are just
the Lagrange multipliers that enforce this condition.
}
\b{Thus we found a variational principle, but indirectly in terms of the
function \m{\rho(x)} rather than the moments \m{G_n} themselves. The entropy
could not be expressed explicitly in terms of the moments. Indeed, in a sense,
this is impossible:
}

\s{The entropy cannot be expressed as a  formal power series in \m{G_n}.
}
\b{This is surprising  since there appears to be a linear relation
between \m{\rho(x)} and \m{G_n}, since \m{G_n=\int x^n\rho(x)dx}; also the
entropy  is a quadratic function of \m{\rho(x)}. So one might think that entropy
is  a quadratic function of the \m{G_n} as well. But if we try to compute this
function we will get a divergent answer. Indeed, we claim that even if we dont
require the series to converge, the entropy cannot be expressed as a power
series in \m{G_n}.
}

\s{By thinking in terms of the change of variables that bring the probability
distribution we seek to a standard one, we can find an explicit formula for
entropy.
}
\b{
Since we are interested in
polynomial actions \m{S(A)}, which are modifications of the quadratic action
\m{\half A^2}, the right choice of this reference distribution is
 the Wigner distribution
\beq
    \rho_0(x)={1\over 2\pi}\surd[4-x^2]\theta(|x|<2).
\eeq
}
\b{There should thus be a change of variable \m{\phi(x)} such that
\beq
    G_k=\int x^k\rho(x)dx=\int \phi(x)^k\rho_0(x)dx;
\eeq
in other words,
\beq
    \rho(\phi(x))\phi'(x)=\rho_0(x).
\eeq
}
\b{
Then we get
\beq
\Omega(\phi)=\int S(\phi(x)) \rho_0(x)dx+\int
\log\left[{\phi(x)-\phi(y)\over x-y}\right]\rho_0(x)dx\rho_0(y)dy.
\eeq
We have dropped a constant term-the entropy of the reference
distribution itself. Also it will be convenient to choose the constant of
integration such that \m{\phi(0)=0}.
}
\b{Now we can regard the diffeomorphism as parametrized by its Taylor
coefficients
\beq
    \phi(x)=\sum_{n=1}^\infty \phi_nx^n, \phi_1>0.
\eeq } \b{Although we cannot express the entropy  in terms of the
moments \m{G_n} themselves, we will be able to express both the
entropy and the moments in terms of the parameters \m{{ \phi}_n}.
Thus we have a `parametric form' of the variational problem. It is
this parametric form that we can extend to the case of
multi-matrix models. } \b{Indeed, \beq {
\phi}^k(x)=\sum_{n=1}^\infty x^n\sum_{l_1+l_2+\cdots l_k=n} {
\phi}_{l_1}\cdots { \phi}_{l_k} \eeq so that \beq G_k=
\sum_{n=1}^\infty
 \Gamma_n\sum_{l_1+l_2+\cdots l_k=n}{\phi}_{l_1}
 \cdots {\phi}_{l_k}.
\eeq } 

\b{It is convenient to  factor out the linear
transformation \m{\phi(x)=\phi_1[x+{\tilde \phi}(x)]},where
\m{{\tilde \phi}(x)=\sum_{k=2}^\infty {\tilde \phi}_kx^k}, with
\m{{\tilde \phi}_k=\phi_k[\phi_1]^{-1}}.Then \beqs
\log\left[{\phi(x)-\phi(y)\over x-y}\right]&=&\log\phi_1+
\log\left[1+\sum_{m=2}^\infty {\tilde \phi}_m{x^m-y^m\over
x-y}\right]. \eeqs Using \beq {x^m-y^m\over x-y}=\sum_{k+1+l=m;\
k,l\geq 0}x^ky^l \eeq and expanding the logarithm we get \beqs
    \lefteqn{\log\left[{{ \phi}(x)-{ \phi}(y)\over x-y}\right]=
    \log { \phi}_1} \cr
    & & +\sum_{n=1}^{\infty} {(-1)^{n+1} \over n}
    \sum_{k_1,l_1,\cdots k_n,l_n}
    {\tilde \phi}_{k_1+1+l_1}\cdots {\tilde \phi}_{k_n+1+l_n}x^{k_1+\cdots k_n}
    y^{l_1+\cdots l_n}
\eeqs } \b{It follows then that \beqs \Omega(\phi)&=&
\sum_{k,n=1}^\infty S_k\Gamma_n \sum_{l_1+l_2+\cdots l_k=n} {
\phi}_{l_1}\cdots {\phi}_{l_k}
            +\log { \phi}_1 \cr
    & & +\sum_{n=1}^{\infty} {(-1)^{n+1} \over n}\sum_{k_1,l_1,\cdots k_n,l_n}
    {\tilde \phi}_{k_1+1+l_1}\cdots {\tilde \phi}_{k_n+1+l_n}
    \Gamma_{k_1+\cdots k_n}\Gamma_{l_1+\cdots l_n}
\eeqs
}
\b{While this formula may not be particularly transparent, it does  accomplish
our goal of finding a variational principle that determines the moments.
The parameters \m{{\phi}_k} characterize the probability distribution of the
eigenvalues: they determine the moments \m{G_n} by the above series. By
extremizing the action \m{\Omega} as a function of these \m{{ \phi}_k}, we can then
determine the moments. We will be able to generalize this version of the action
principle to multi-matrix models.
}
\b{In practice we would choose some simple function \m{{ \phi}(x)} such as a
polynomial to get an approximate solution to this variational problem. Since all
one-matrix models are exactly solvable, we can use them to test the accuracy of
our variational approximation.
}

\subsection{Explicit Variational Calculations}

Consider the quartic one matrix model. Its exact solution in the
large $N$ limit is known from the work of Brezin et. al.
\cite{planarsaclay}:

\beqs
    Z(g) &=& \int dA e^{N \tr [-\half A^2 - g A^4]} \cr
    E_{exact}(g) &=& -\lim_{N \to \infty} {1 \over N^2} \log{Z(g) \over Z(0)}
\eeqs

\noindent The gaussian with unit covariance is our reference
action. Choose as a variational ansatz the linear change of
variable $\phi(x) = \phi_1 x$, which merely scales the Wigner
distribution. The $\phi_1$ that maximizes $\Omega$ represents the
Wigner distribution that best approximates the quartic matrix
model.

\beq
    \Omega(\phi_1) = \log \phi_1 - \half G_2 - g G_4
\eeq

\noindent Here $G_{2k} = \phi_1^{2k} \Gamma_k$. Letting $\alpha =
\phi_1^2$, $\Omega(\alpha) = \half \log \alpha - {\alpha \over 2}
- 2g\alpha^2$ is bounded above only for $g \geq 0$. It's maximum
occurs at $\alpha(g) = {-1 + \sqrt{1+32g} \over 16 g}$. Notice
that $\alpha$ is determined by a non-linear equation. This is
reminiscent of mean field theory; we will sometimes refer to the
gaussian ansatz as mean field theory. Our variational estimates
are:

\beqs
    E(g) &=& -\half \log{{-1 + \sqrt{1+32g} \over 16 g}} \cr
    G_{2k}(g) &=& ({-1 + \sqrt{1+32g} \over 16 g})^k C_{k}.
\eeqs

\noindent The exact results from \cite{planarsaclay} are:

\beqs
    E_{ex}(g) &=& {1 \over 24} (a^2(g) - 1)(9-a^2(g)) - \half
                    \log{(a^2(g))} \cr
    G_{ex}^{2k}(g) &=& {(2k)! \over k! (k+2)!} a^{2k}(g) [2k + 2
                        -ka^2(g)].
\eeqs

\noindent where $a^2(g) = {1 \over 24 g} [-1 + \sqrt{1+48 g}]$. In
both cases, the vacuum energy is analytic at $g = 0$ with a square
root branch point at a negative critical coupling. The mean field
critical coupling $g_c^{MF} = -{1 \over 32}$ is $50\%$ more than
the exact value $g_c^{ex} = -{1 \over 48}$.

The distribution of eigenvalues of the best gaussian approximation
is given by $\rho_g(x) = \phi_1^{-1} \rho_0(\phi_1^{-1} x)$ where
$\rho_0(x) = {1 \over 2\pi} \sqrt{4 - x^2}, ~ |x| \leq 2$ is the
standard Wigner distribution. The exact distribution

\beq
    \rho_{ex}(x,g) = {1 \over \pi} (\half + 4 g a^2(g) + 2 g x^2)
    \sqrt{4 a^2(g) - x^2}, ~ |x| \leq 2 a(g).
\eeq

\noindent is compared with the best gaussian approximation in
figure \ref{fig-1-mat-1-loop-rho}. The latter does not capture the
bimodal property of the former.

The vacuum energy estimate starts out for small $g$, being twice
as big as the exact value. But the estimate improves and becomes
exact as $g \to \infty$. Meanwhile, the estimate for $G_2(g)$ is
within $10 \%$ of its exact value for all $g$. $G_{2k}, k \geq 2$
for the gaussian ansatz do not have any new information. However,
the higher cumulants vanish for this ansatz.

We see that a gaussian ansatz is a reasonable first approximation,
and is not restricted to small values of the coupling $g$. To
improve on this, get a non-trivial estimate for the higher
cumulants and capture the bimodal distribution of eigenvalues, we
need to make a non-linear change of variable.

\subsection{Non-linear Variational Change of Variables}

The simplest non-linear ansatz for the quartic model is a cubic
polynomial: $\phi(x) = \phi_1 x + \phi_3 x^3$. A quadratic ansatz
will not lower the energy since $S(A)$ is even. Our reference
distribution is still the standard Wigner distribution.
$\phi_{1,3}$ are determined by the condition that

\beq
    \Omega[\phi] = <\log|{\phi(x) - \phi(y) \over x-y}|>_0
        - \half <\phi^2(x)> - g <\phi^4(x)>_0
\eeq

\noindent be a maximum. Considering the success of the linear
change of variable, we expect the deviations of $\phi_{1,3}$ from
their mean field values $(\sqrt{\alpha},0)$ to be small,
irrespective of $g$. Within this approximation, we get (with
$\alpha = {-1 + \sqrt{1+32g} \over 16 g}$)

\beqs
    \phi_1 &=& \sqrt{\alpha} - {\sqrt{\alpha}(-3 + 2 \alpha + (1-32 g) \alpha^2
        + 48 g \alpha^3 + 144 g^2 \alpha^4) \over 3 + 4\alpha +
        (1+96g)\alpha^2 + 48g \alpha^3 + 432g^2 \alpha^4} \cr \cr
    \phi_3 &=& {8 g \alpha^{5 \over 2} (-2 + \alpha) \over 3
        + 4\alpha + (1+96g)\alpha^2 + 48g \alpha^3 + 432g^2 \alpha^4}.
\eeqs

\noindent from which we calculate the variational greens functions
and vacuum energy. The procedure we have used to obtain
$\phi_{1,3}$ can be thought of as a $1$ loop calculation around
mean field theory. Comparing with the exact results of
\cite{planarsaclay}, we find the following qualitative
improvements over the mean field ansatz.

In addition to the mean field branch cut from $-\infty$ to
$g_c^{MF}$, the vacuum energy now has a double pole at $g_c^{MF} <
g_c^{1-\rm{loop}} = {-346 - 25 \sqrt{22} \over 15138} < g_c^{ex}$.
We can understand this double pole as a sort of Pad$\acute{\rm e}$
approximation to a branch cut that runs all the way up to
$g_c^{ex}$. The vacuum energy variational estimate is lowered for
all $g$.

\begin{figure}
\centerline{\epsfxsize=6.truecm\epsfbox{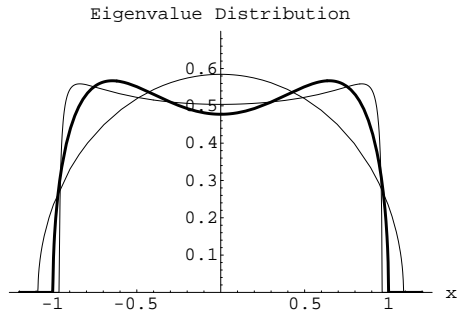}}
\caption{ Eigenvalue Distribution. Dark curve is exact, semicircle
is mean field and bi-modal light curve is cubic ansatz at
$1$-loop.} \label{fig-1-mat-1-loop-rho}
\end{figure}

Figure \ref{fig-1-mat-1-loop-rho} demonstrates that the cubic
ansatz is able to capture the bimodal nature of the exact
eigenvalue distribution. If $\chi(x) = \phi^{-1}(x)$, then
$\rho(x) = \rho_0(\chi(x)) \chi'(x)$, where $\rho_0(x) = {1 \over
2\pi} \sqrt{4-x^2}, ~ |x| \leq 2$.

The greens functions $G_2, G_4$ are now within a percent of their
exact values, for all $g$. More significantly, the connected
4-point function $G_4^c = G_4 - 2 (G_2)^2$ which vanished for the
gaussian ansatz, is non-trivial, and within $10\%$ of its exact
value, across all values of $g$.

\subsection{ Formal Power Series in One Variable}

\s{ Given a sequence of complex numbers
\m{(a_0,a_1,a_2,\cdots)},  with only a
finite number of non-zero entries, we have a polynomial with these
numbers as coefficients\cite{cartan}:
\beq
a(z)=\sum_{n=0}^\infty  a_n z^n.
\eeq
}
\b{ Note that all the information in a polynomial is in its
coefficients: the  variable \m{z} is just a book-keeping device. In
fact we could have defined a polynomial as a sequence of complex
numbers  \m{(a_0,a_1,\cdots)} with  a finite number of non-zero
elements. The addition multiplication and division of polynomials can
be expressed directly in terms of these coefficients:
\beq
[a+b]_n=a_n+b_n,\quad [ab]_n=\sum_{\matrix{k+l=n}}a_kb_l,\quad [Da]_n=(n+1)a_{n+1}.
\eeq
}

\s{ A {\Em formal power series} \m{(a_0,a_1,\cdots)} is  a
 sequence of complex numbers, with possibly an infinite number of non-zero
terms. We  define the sum,  product and derivative as for polynomials above:
\beq
    [a+b]_n=a_n+b_n,\quad
[ab]_n=\sum_{\matrix{k+l=n}}a_kb_l,\quad [Da]_n=(n+1)a_{n+1}.
\eeq
The set of formal power series is a ring, indeed even an integral
domain. ( The proof is the same as above for polynomials.) The
 opration \m{D} is a derivation on this ring.
The ring  of formal power series  is  often denoted by \m{C[[z]]}.
}
\b{ The idea is that such a sequence can be thought of as the coefficients of
 a series
 \m{\sum_{n=0}^\infty a_nz^n}; the sum and product postulated
are what you would get
from this interpretation. However, the series may not make converge if \m{z}
is thought of as a complex number: hence the name {\Em formal} power series.
}

\s{ The   {\Em composition} \m{a\circ b} is well-defined whenever  \m{b_0=0}:
\beq
[a\circ b]_n= \sum_{k=0}^\infty a_k \sum_{l_1+\cdots l_k=n} b_{l_1}b_{l_2}\cdots  b_{l_k}.
\eeq
The point is that, for each \m{n} there are only a finite number of
such \m{l}'s so that the series on the rhs is really a finite
series.
}
\b{ In terms of series, this means we
substitute one series into the other:
\beq
    a\circ b(z)=a(b(z)).
\eeq
}

\subsection{ The Group of automorphisms}

\s{ The set of formal power series
\beq
    {\cal G}=\{\phi(z)=\sum_{0}^\infty \phi_nz^n|\phi_0=0;\phi_1\neq 0\}
\eeq
is a group under  composition:
 the group of  {\Em automorphisms}. The group  law  is
\beq
[{\tilde \phi}\circ \phi]_n=\sum_{k=1}^n{\tilde \phi}_k\sum_{l_1+l_2\cdots+
l_k=n}\phi_{l_1}\cdots \phi_{l_k}
\eeq The inverse of \m{\phi} (say \m{{\tilde \phi}})
is determined by the  recursion relations of Lagrange:
\beq
    {\tilde \phi}_1\phi_1=1,\quad {\tilde \phi}_n=-{1\over
\phi_1^n}\sum_{k=1}^{n-1}{\tilde \phi}_k\sum_{l_1+\cdots l_k=n}\phi_{l_1}\cdots \phi_{l_k}.
\eeq
}

\s{  \m{\cal G} is a topological group with respect to the
ultrametric topology. It can be thought of as a Lie
group, the coefficients \m{\phi_n} being the co-ordinates. The group
multiplication law  can now be studied in the case where the left or
the right element is infinitesimal, leading to two sets of vector
fields on the group manifold. For example, if \m{{\tilde \phi}(x)=x+\eps x^{k+1}},
the change it would induce on the co-ordinates of \m{\phi} is
\beq
    [{\cal L}_k\phi]_n=
\sum_{l_1+l_2\cdots l_{k+1}=n}\phi_{l_1}\cdots \phi_{l_{k+1}}
\eeq
or equivalently,
\beq
    {\cal L}_k\phi(x)=\phi(x)^{k+1}, \; {\rm for}\; k=0,1,2\cdots.
\eeq
By  choosing  \m{{\tilde \phi}(x)=x+\eps x^{k+1}} in \m{\phi\circ {\tilde \phi}} we get the
infinitesimal right action:
\beq
    {\cal R}_k\phi(x)=x^{k+1}D\phi(x), \; {\rm for}\; k=0,1,2\cdots.
\eeq
Both sets  satisfy the commutation relations of the Lie algebra
\m{\un{\cal G}}:
\beq
    [{\cal L}_m,{\cal L}_n]=(n-m){\cal L}_{m+n},\quad
[{\cal R}_m,{\cal R}_n]=(n-m){\cal R}_{m+n}.
\eeq
This Lie algebra is also called the {\Em Virasoro} algebra by some
physicists and the {\Em Witt} algebra by some mathematicians.
}

\s{There is a representation of this Lie algebra on the space of
formal power series:
\beq
    L_na=x^{n+1}Da
\eeq
}

\subsection{Cohomology of \m{\cal G}}

\s{ Now let \m{V} be the space of formal power series with real
coefficients. Then \m{\cal G}, the group of automorphims has a
representation on \m{V}:
\beqs
{\cal G}&=&\{\phi:Z_+\to R|\phi_0=0,\phi_1>0\},\cr
 V&=&\{a:Z_+\to
R\},\quad
\phi^*a(x)=a(\phi^{-1}(x)) .
\eeqs
}

\b{Now, \m{\log[\phi(x)/x]} is a power series in \m{x} because
 \m{\phi(x)/x} is a formal power series with positive  constant term:
 \m{[\phi(x)/x]_0=
\phi_1>0}.
}
\b{We see easily that \m{c(\phi,x)=\log[\phi(x)/x)]} is a \m{1}-cocycle of \m{\cal
G} twisted by the representation \m{ V}:
\beqs
    c(\phi_1 \phi_2,x)&=&\log\left[{\phi_1(\phi_2(x))\over
x}\right],\cr
 &=&\log\left[{\phi_1(\phi_2(x))\over \phi_2(x)}\right]+
 \log\left[{\phi_2(x)\over x}\right]\cr
 &=&c(\phi_1\phi_2,\phi_2(x))+c(\phi_2,x) .
\eeqs
}
\b{
Of course,  neither \m{\log \phi(x)} nor \m{\log x} are  power series in
\m{x}. So this cocycle is non-trivial.
}

\s{ The space of formal power series in two commuting variables (
\m{{\rm Sym}^2\ V}) also carries a representation of \m{\cal G}. 
We again have a non-trivial \m{1}-cocycle\footnote{The formula
\beq
    {x^m-y^m\over x-y}=\sum_{ k=1}^{m-1} x^ky^{m-k-1}
\eeq
can be used to show that \m{c(\phi,x,y)}
 is  a formal power series in \m{x} and \m{y}.} on this representation:
\beq
c(\phi,x,y)=\log\left[{\phi(x)-\phi(y)\over x-y}\right].
\eeq
}
\b{We recognize this as the entropy of the single matrix model. The same
argument shows that this is a non-trivial cocycle of \m{\cal G}.
}

\s{Now we understand that the entropy of the single matrix models has the
mathematical meaning of a non-trivial 1-cocycle of the group of automorphisms.
 It explains why we could not express the entropy as a function of the moments.
 This points also to a solution to the difficulty: we must think in terms of the
 automorphism \m{\phi} rather than the moments as parametrizing the probability
 distribution.
}
\section{ Appendix: Formula for Cocycle}

\s{We will now  get an explicit formula for \m{\sigma(\tilde\phi,A)}.
}
\b{ The Jacobian matrix of \m{\tilde \phi} is obtained by differentiating the
series \m{{\tilde \phi}(A)_i=A_i +\sum_{n=2}^\infty{\tilde \phi}_i^{i_1\cdots i_n}
A_{i_1}\cdots A_{i_n}}:
\beq
    J^{a\ jd}_{ib\ c}(A)&=&{\pdr \tilde \phi^{a}_{ib}(A)\over\pdr A^{c}_{jd}}\cr
            &=&\delta^j_i\delta^a_c\delta^d_b+\cr
	    & & 
\sum_{m+n\geq 1}\tilde\phi_i^{i_1\cdots i_m j j_1\cdots j_n}
\left[A_{i_1}\cdots A_{i_m}\right]^a_c
\left[A_{j_1}\cdots A_{j_n}\right]^d_b\cr
&:=& \delta^j_i\delta^a_c\delta^d_b  +K^{a\ jd}_{ib\ c}(A).
\eeq
}

\b{
If we suppress the color indices \m{a,b,c,d},
\beq
    J_i^j(A)=\delta_i^j 1\otimes 1 +\tilde\phi_i^{IjJ}A_I\otimes
A_J:=\delta_i^j 1\otimes 1+K_i^j(A) \eeq } \b{We can now compute
\beq\nonumber \lefteqn{{1\over N^2}\tr K^n(A)={1\over N^2}K^{a_1\
i_2b_2}_{i_1b_1\ a_2} K^{a_2\ i_3b_3}_{i_2b_2\ a_2}\cdots K^{a_n\
i_1b_1}_{i_nb_n\ a_1}}\cr
&=&\tilde\phi_{i_1}^{K_1i_2L_1}\tilde\phi_{i_2}^{K_2i_3L_2}\cdots
\tilde\phi_{i_n}^{K_ni_1L_n}\cr & & {1\over
N}\left[A_{K_1}\right]^{a_1}_{a_2}\cdots
\left[A_{K_n}\right]^{a_n}_{a_1} {1\over
N}\left[A_{L_1}\right]^{b_2}_{b_1}\cdots
\left[A_{L_n}\right]^{b_1}_{b_n}\cr
&=&\tilde\phi_{i_1}^{K_1i_2L_1}\tilde\phi_{i_2}^{K_2i_3L_2}\cdots
\tilde\phi_{i_n}^{K_ni_1L_n}\Phi_{K_1\cdots K_n}\Phi_{L_n\cdots
L_1}. \eeq } 

\b{ Thus, \beq\nonumber
\sigma(\tilde\phi,A)&=&{1\over
N^2}\log\det\left[1+K(A)\right]\cr
    &=& \sum_{n=1}^\infty {(-1)^{n+1}\over n}{1\over N^2}\tr K^n\cr
&=&\sum_{n=1}^{\infty}{(-1)^{n+1}\over n}
\tilde\phi_{i_1}^{K_1i_2L_1}\tilde\phi_{i_2}^{K_2i_3L_2}\cdots
\tilde\phi_{i_n}^{K_ni_1L_n}\Phi_{K_1\cdots K_n}\Phi_{L_n\cdots
L_1}. 
\eeq 
} 
\b{
This is the formula we presented earlier.
}


\begin{thebibliography}{10}

 \bibitem{thooft}  G. 't Hooft Nucl.Phys. B75, 461 (1974); 
{\it Under the Spell of
 the Gauge Principle}, (World Scientific,Singapore (1994)). 

 \bibitem{wittenN} E. Witten, Nucl.Phys. B160, 57 (1979).

 \bibitem{makeenko} Y.M.  Makeenko and A. A. Migdal,
 Phys.Lett. B88, 135 (1979), Erratum-ibid.B89, 437 (1980);
 Nucl.Phys. B188, 269 (1981).
 
 \bibitem{2dqhd} S. G. Rajeev, Int.J.Mod.Phys. A9, 5583 (1994).
 
 \bibitem{quarkparton} G. Krishnaswami and S. G. Rajeev, 
 Phys.Lett.B441, 429 (1998);
 V. John, G.S. Krishnaswami and  S.G. Rajeev Phys.Lett.B487, 125 (2000);
 Phys.Lett. B492, 63 (2000); 
 {\it Theoretical  High Energy Physics: MRST 2000} 
 ed. C.R. Hagen. (Amer. Inst. Phys., Melville, N.Y., 2000);
 S. G. Rajeev, Nucl.Phys.Proc.Suppl.96 487 (2001); hep-th/9905072.

\bibitem{rajeevturgut}S.G. Rajeev and  O.T. Turgut, 
Comm. Math. Phys.192, 493 (1998);
  J. Math. Phys. 37, 637 (1996); 
  Int. J. Mod. Phys. A 10, 2479 (1995).

\bibitem{mrst2001} L. Akant, G.S. Krishnaswami and S.G. Rajeev, in {\it
Theoretical  High Energy Physics: MRST 2001} 
 eds. V.Elias et. al. (Amer. Inst. Phys., Melville, N.Y. 267, 2001).

\bibitem{wigner}E. Wigner, Proc. Cambridge Philos. Soc. 47, 790 (1951);
reprinted in C. E. Porter, {\it Statistical Theories of Spectra: Fluctuations}
 (Academic Press, New York, 1965).

\bibitem{dyson} F.J. Dyson, J. Math. Phys. 3, 140,157,166 (1962).

\bibitem{mehta} M. L. Mehta {\it Random Matrices}, (Academic Press, 
New York, 1991) 2nd ed.

\bibitem{cvitanovic} P.  Cvitanovic Phys.Lett. B99, 49 (1981);
 P. Cvitanovic, P.G. Lauwers and  P.N. Scharbach, Nucl.Phys. B186, 
 165 (1981).

\bibitem{douglas}
M.~R.~Douglas and M.~Li,
Phys.\ Lett.\ B {\bf 348}, 360 (1995).
M.~R.~Douglas,
Fields Institute Communications Vol 12, (1995)

\bibitem{gross-gopakumar}
R.~Gopakumar and D.~J.~Gross,
Nucl.\ Phys.\ B {\bf 451}, 379 (1995)

\bibitem{voiculescu} D. V.  Voiculescu, K. J. Dykema and A. Nica
  {\it Free Random Variables},  (American Mathematical Society, Providence 
USA, 1992); D. V.  Voiculescu, Invent. Math. 118, 411 (1994); 132, 189 (1998).

\bibitem{guhr} T. Guhr, A. Mueller-Groeling and H. A. Weidenmuller, Phys. Rep.
299, 189 (1998).

\bibitem{senreview} A. Sen,Nucl.Phys.Proc.Suppl. 94, 35 (2001).

\bibitem{ambjorn} J. Ambjorn, B. Durhuus, T. Jonsson,
{\it  Quantum Geometry}, (Cambridge Univ. Press, 1997); 
J. Ambjorn, J. Jurkiewicz, R. Loll,
 Phys.Rev. D64, 044011 (2001).

\bibitem{akhiezer} N. I. Akhiezer, {\it Classical Moment Problem}, 
(Hafner, New York, 1965).

\bibitem{ambjornmakeenko}J. Ambjorn, J. Jurkiewicz and Y. M. Makeenko, Phys.
Lett. B251, 517 (1990).

\bibitem{evens}L. Evens, {\it The Cohomology of Groups}, (Oxford, 1991).

\bibitem{ferrettirajeev} G. Ferretti and S. G. Rajeev, Phys. Rev. Lett. 69
 2033 (1992).

\bibitem{planarsaclay} E. Brezin, C. Itzykson, G. Parisi and  J.B. Zuber,
 Comm.Math.Phys.59, 35 (1978).

\bibitem{brezinzee} E. Brezin and A. Zee, Nucl. Phys. B402, 613 (1993).

\bibitem{mehta-two-matrix}
M.~L.~Mehta,
Commun.\ Math.\ Phys.\   79, 327 (1978).

\bibitem{Kazakov-rand-ising}
V.~A.~Kazakov,
Phys.\ Lett.\ A  119, 140 (1986).

\bibitem{Staudacher-two-mat}
M.~Staudacher,
Phys.\ Lett.\ B {\bf 305},  332 (1993).

\bibitem{kawai-et-al}
N.~Ishibashi, H.~Kawai, Y.~Kitazawa and A.~Tsuchiya,
Nucl.\ Phys.\ B {\bf 498}, 467 (1997).

\bibitem{cartan} H. Cartan, {\it Elementary 
Theory of Analytic Functions of One or Several Complex Variables }
(Hermann, Paris, 1973).

\end{thebibliography}
\end{document}